\documentstyle[fleqn,epsfig,run2col]{article}

\newcommand{\AmS}{{\protect\the\textfont2
  A\kern-.1667em\lower.5ex\hbox{M}\kern-.125emS}}
\def\GeV{{\rm GeV}}

\title{\vspace*{-4\baselineskip}
\hfill {\rm UR-1599}\\
\hfill {\rm ER/40685/944}\\
\hfill {\rm December 1999}\\
\vspace*{\baselineskip}
BFKL Monte Carlo for  Dijet Production at Hadron 
Colliders\thanks{Presented at the Fermilab Run II Workshop, QCD and 
Weak Boson Physics, June 3--4, 1999.}}

\author{Lynne H.~Orr\address{Department of Physics and Astronomy, 
        University of Rochester \\ 
        Rochester NY 14627-0171}%
        \thanks{Work supported in part by the U.S. Department of Energy,
under grant DE-FG02-91ER40685 and by the U.S. National Science Foundation, 
under grants PHY-9600155 and PHY-9400059.}
        and 
        W.J.~Stirling\address{Departments of Physics and Mathematical Sciences, 
        University of Durham, \\
        Durham DH1 3LE, UK}}
       
\begin{document}

\begin{abstract}
The production of jet pairs at large rapidity difference at hadron 
colliders is potentially sensitive to BFKL physics.  We present the 
results of a BFKL Monte Carlo calculation of dijets at the Tevatron.
The Monte Carlo incorporates kinematic effects that are absent in analytic
BFKL calculations; these effects significantly modify the behavior 
of dijet cross sections.
\end{abstract}

\maketitle

\section{MONTE CARLO APPROACH TO BFKL}

Fixed-order QCD perturbation theory fails in  
some asymptotic regimes  where  large logarithms multiply
the coupling constant.  In those regimes resummation of the perturbation 
series  to all orders is necessary to describe many high-energy processes.
The Balitsky-Fadin-Kuraev-Lipatov (BFKL) equation~\cite{bfkl} performs such a
resummation for virtual and real soft gluon emissions in  such processes as  
dijet production at large rapidity difference in hadron-hadron collisions. 
BFKL resummation gives~\cite{muenav} a
subprocess cross section that increases with rapidity difference as
$\hat\sigma\sim\exp(\lambda \Delta)$,
where $\Delta$ is the rapidity difference of the two jets with comparable
transverse momenta $p_{T1}$ and $p_{T2}$.

Experimental studies of these processes have recently begun at the  
Tevatron $p \bar p$ and HERA $ep$ colliders.  
Tests so far have been inconclusive;  the data tend to lie between
fixed-order QCD and analytic BFKL predictions.  However the 
applicability of analytic BFKL solutions is limited by the
fact that they implicitly contain integrations over arbitrary numbers
of emitted gluons with arbitrarily large transverse momentum:  there
are no kinematic constraints included.  Furthermore, 
the implicit sum 
over emitted gluons leaves only leading-order kinematics, including 
only the momenta of the `external' particles.
The absence of kinematic constraints and energy-momentum conservation cannot,
of course, be reproduced in experiments.  While the effects of such constraints
are in principle sub-leading, in fact they can be  substantial and
 should be included in
predictions to be compared with experimental results.

The solution is to unfold the implicit sum over gluons 
and to implement the result in a Monte Carlo
event generator~\cite{os,schmidt}.  This is achieved as follows.
The BFKL equation contains separate integrals over  real and virtual 
emitted gluons.  We can reorganize the equation by combining the 
`unresolved' real emissions --- those with transverse momenta
below some minimum value (chosen to be small
compared to the momentum threshold for measured
jets) --- with the virtual emissions.  Schematically,
we have 
\begin{equation}
\int_{virtual} + \int_{real} = \int_{virtual+real, unres.} +
\int_{real, res.}
\end{equation}
We  perform
the integration over virtual and unresolved real
emissions  analytically.  The integral containing the 
resolvable real emissions is left explicit. 

We then solve by iteration, 
and we obtain a differential cross section
that contains a sum over emitted gluons along with 
the appropriate phase space factors.  In addition, we obtain
an overall form factor due
to virtual and unresolved emissions. The subprocess cross section is
\begin{equation}
d\hat\sigma=d\hat\sigma_0\times\sum_{n\ge 0} f_{n}
\end{equation}
where $f_{n}$ is the iterated solution for $n$ real gluons emitted and
contains the overall form factor.
It is then straightforward to implement the result in a Monte Carlo
event generator.   Because emitted real (resolved) gluons appear explicitly, 
conservation of momentum and energy, as well as
evaluation of parton distributions, 
is based on exact kinematics
for each event.  In addition, we include the running of the strong
coupling constant.  See~\cite{os} for further details.

\section{DIJET PRODUCTION AT HADRON COLLIDERS}

\begin{figure*}[t]
\psfig{figure=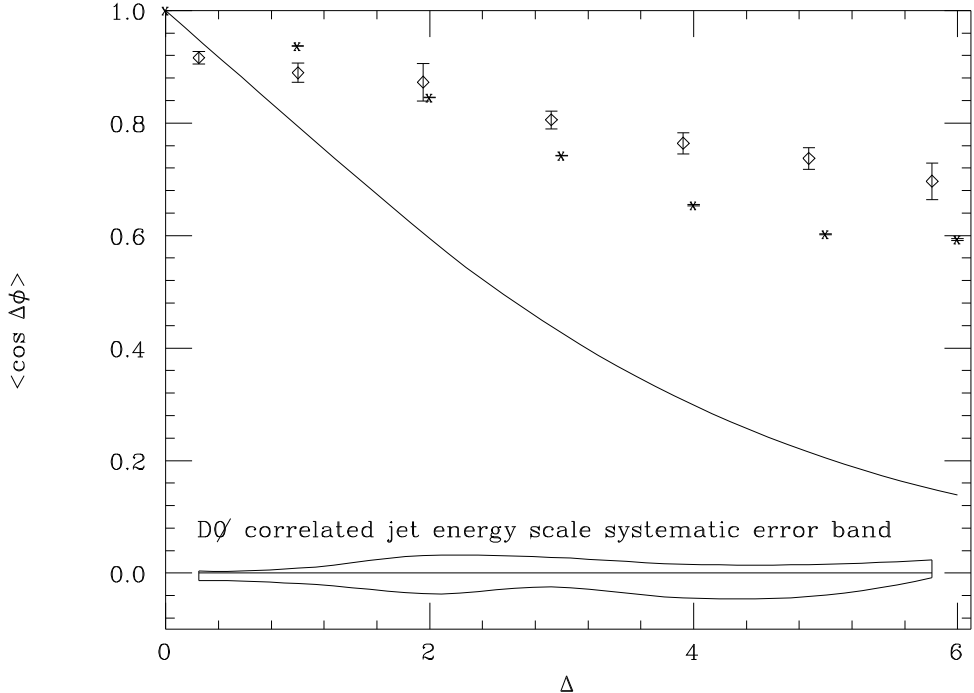,width=16.0cm}
\vskip -1.25cm
\caption[]{The azimuthal angle decorrelation in dijet production at the 
Tevatron 
as a function of dijet rapidity difference $\Delta$, for 
jet transverse momentum $p_T>20\ {\rm GeV}$.  
The analytic BFKL solution is shown as a solid curve 
and a preliminary D$\emptyset$\ measurement~\cite{dzeropl} is shown
as diamonds.  Error bars represent statistical and 
uncorrelated systematic errors;  correlated jet energy scale systematics
are shown as an error band.  \label{fig:decor}}
\end{figure*}

At hadron colliders, the BFKL increase in the dijet subprocess cross section 
with rapidity difference $\Delta$ is unfortunately washed out by the falling
parton distribution functions (pdfs).  As a result, the BFKL prediction for 
the total cross section is simply a less steep falloff than obtained in 
fixed-order QCD, and tests of this prediction are sensitive to pdf
uncertainties.  A more robust pediction is obtained by noting that 
the emitted gluons  give rise to
a decorrelation in azimuth between the two leading jets.\cite{many,os}  This 
decorrelation becomes stronger as  $\Delta$ increases
and more gluons are emitted.  In lowest order in QCD, in contrast, the jets
are back-to-back in azimuth and the (subprocess) cross section is 
constant, independent
of $\Delta$.

This azimuthal decorrelation is illustrated in Figure~\ref{fig:decor}
for dijet production at the Tevatron $p\bar p$ collider~\cite{os}, with 
center of mass energy 1.8 TeV  and jet transverse momentum $p_T>20\ {\rm GeV}$.
The azimuthal angle difference $\Delta\phi$ is defined such that 
$\cos\Delta\phi=1$ for back-to-back jets.
The solid line shows the analytic BFKl prediction.  The BFKL Monte Carlo
prediction is shown as crosses.  We see that the kinematic constraints
result in a weaker decorrelation due to suppression of 
emitted gluons, and we obtain improved  agreement with
preliminary measurements by the D$\emptyset$\ collaboration~\cite{dzeropl}, 
shown as diamonds in the figure.
 
\begin{figure}[t]
\psfig{figure=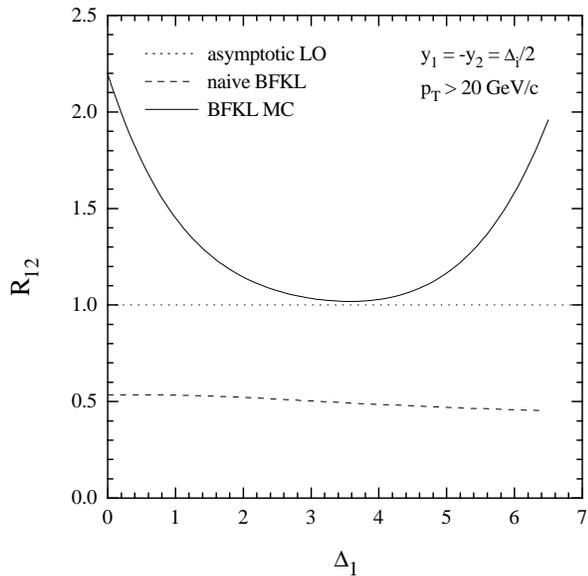,width=9.cm}
\vskip -1.25cm
\caption[]{
The  ratio $R_{12}$ of the dijet cross sections at the two collider energies
$\sqrt s_1 = 630\; \GeV$ and $\sqrt s_2 = 1800 \; \GeV$, as defined in the text.
The curves are: (i) the  BFKL MC predictions (solid curve),  
(ii) the `naive' BFKL prediction (dashed curve), and 
(iii)  the
asymptotic QCD leading-order prediction (dotted curve)  $R_{12}=1$.
}
\label{fig:ronetwobfkl}
\end{figure}

In addition to studying the azimuthal decorrelation, one can 
look for the BFKL rise in  dijet cross section with
rapidity difference by considering  ratios of cross sections
at different center of mass energies at fixed $\Delta$.
The idea is to cancel the pdf dependence, leaving the pure
BFKL effect.  This turns out to be rather tricky~\cite{osratio},
because the desired  cancellations occur only at lowest
order. 
Therefore we consider the ratio 
\begin{equation}
R_{12} ={d\sigma(\sqrt s_1,\Delta_1) \over  d\sigma(\sqrt s_2,\Delta_2)}
\end{equation}
with $\Delta_2$ defined such that $R_{12} = 1$ in QCD lowest-order.  
the result is shown in Figure~\ref{fig:ronetwobfkl}, and we see
that the kinematic constraints strongly affect the predicted
behavior, not only quantitatively but sometimes 
qualitatively as well.  More details can be found in \cite{osratio}.

\section{CONCLUSIONS}

In summary, we have developed a BFKL Monte Carlo event generator that 
allows us to include the subleading effects such as kinematic constraints
and running of $\alpha_s$.  We have applied this Monte Carlo to 
dijet production at large rapidity separation at the Tevatron.  
We found that kinematic constraints, though nominally 
subleading, can be very important.  In particular they lead to suppression
of gluon emission, which in turn suppresses some of the behavior that is 
considered to be characteristic of BFKL physics.  It is clear therefore
that reliable BFKL tests  can only be performed using predictions
that incorporate kinematic constraints.

\end{document}